\documentclass[twocolumn,showpacs,amsmath,amssymb,prb]{revtex4-1}
\usepackage{graphicx}% Include figure files
\usepackage{dcolumn}% Align table columns on decimal point
\usepackage{bm}% bold math
\usepackage[abs]{overpic}
\usepackage{here}
\usepackage{color}
\def\up{\uparrow}
\def\down{\downarrow }
\def\Vec#1{\bm{#1}}

\begin{document}

%\preprint{}

\title{Nuclear magnetic relaxation rates of unconventional superconductivity in doped topological insulators
}

\author{Yuki Nagai}
\affiliation{CCSE, Japan  Atomic Energy Agency, 178-4-4, Wakashiba, Kashiwa, Chiba, 277-0871, Japan}

\author{Yukihiro Ota}
\affiliation{Research Organization for Information Science and Technology (RIST), 1-5-2 Minatojima-minamimachi, Kobe, 650-0047, Japan}

\date{\today}% It is always \today, today,
             %  but any date may be explicitly specified
             
\begin{abstract}
We study the temperature dependence of nuclear magnetic relaxation (NMR) rates to detect a sign of topological superconductivity in doped topological insulators, such as 
 {\it M}($=$Cu,Nb,Sr)$_{x}$Bi$_{2}$Se$_{3}$ and Sn$_{1-x}$In$_{x}$Te. 
The Hebel-Slichter coherence effect below a critical temperature $T_{\rm c}$ 
depends on the superconducting states predicted by a minimal model of doped topological insulators.
In a nodal anisotropic topological state similar to the ABM-phase in $^{3}$He, the NMR rate has a conventional $s$-wave like coherence peak below $T_{\rm c}$. 
In contrast, in a fully-gapped isotropic topological superconducting state, this rate below $T_{\rm c}$ exhibits an anti-peak profile. 
Moreover, in a two-fold in-plane anisotropic topological superconducting state, there is no coherence effect, which is similar to that in a chiral $p$-wave state. 
Thus, we reveal that 
the NMR rates shed light on unconventional superconductivity in doped topological insulators.
\end{abstract}

% PACS, the Physics and Astronomy
                             % Classification Scheme.
%\keywords{Suggested keywords}%Use showkeys class option if keyword
                              %display desired
\maketitle

\section{Introduction}
The recent discovery of topological insulators \cite{Bernevig15122006,PhysRevLett.105.266401,PhysRevB.76.045302,PhysRevLett.98.106803,PhysRevLett.95.146802,Konig02112007,PhysRevLett.105.146801,PhysRevB.75.121306,PhysRevB.81.041309,PhysRevLett.105.136802} leads to a 
number of studies about topological aspects in condensed matter physics\cite{RevModPhys.82.3045}. 
Topological superconductors are also attracted much attention  because of their potential applications for topological quantum computing\cite{PhysRevLett.104.046401}. 
The quest for bulk topological superconductors is an exciting issue in topological material science. 
Both surface and bulk probes are crucial for identifying topological materials. 
The bulk-boundary correspondence indicates that the presence of gapless surface bound states between different topological materials 
is an evidence of a nontrivial topological order. 
Bulk quantities also contain a signature of their topological order. 
A definite example in condensed matter physics is the conductance in the integer quantum Hall systems\cite{Textbook}. 
Quantized behaviors ruled by a topological invariant are observed. 

Doped topological insulators are candidates of 3D time-reversal invariant topological 
superconductors with $Z_2$ invariants\cite{PhysRevLett.105.097001,PhysRevB.81.220504}.
Bi$_{2}$Se$_{3}$ has a superconducting critical temperature $T_{\rm c}$ about 3K with Cu, Sr, and Nb doping\cite{PhysRevLett.104.057001,PhysRevB.92.020506,1.4934590,arXiv:1603.04040}. 
Doped topological crystalline insulator Sn$_{1-x}$In$_{x}$Te also becomes a superconductor with $T_{\rm c} \sim 4$K\cite{PhysRevLett.109.217004,PhysRevB.88.020505}. 
The properties of Cu$_{x}$Bi$_{2}$Se$_{3}$ and Sn$_{1-x}$In$_{x}$Te are studied by different physical probes, including point-contact spectroscopy\cite{PhysRevLett.107.217001,PhysRevB.86.064517,PhysRevLett.109.217004}, scanning tunneling spectroscopy\cite{PhysRevLett.110.117001}, 
the Knight-shift measurement\cite{arXiv:1512.07086}, and the angular-resolved heat capacity measurement\cite{arXiv:1602.08941}. 
The point-contact spectroscopy showed the zero-bias conductance peaks from the Majorana bound states at the surface edges. 
The scanning tunneling spectroscopy, however, indicated a fully gapped feature in the density of states; 
there is no in-gap state, and therefore the superconducting state could be topologically trivial. 
In addition, the Knight-shift and the angular-resolved heat capacity measurements 
on $\mbox{Cu}_x \mbox{Bi}_2 \mbox{Se}_3$ 
showed the presence of two-fold in-plane anisotropy. 
This result indicates a strong anisotropic order parameter since 
the normal-state electronic structure 
has a six-fold in-plane symmetry caused by the crystal structure\cite{JPSJ.82.044704,PhysRevB.86.094507,JPSJ.83.063705,PhysRevB.90.100509}.  
A similar anisotropic feature is also observed by the torque magnetometry measurement  in $\mbox{Nb}_x \mbox{Bi}_2 \mbox{Se}_3$\cite{arXiv:1603.04040}. 
Thus, the doped topological insulators have unconventional properties in their superconducting states, which might be topologically nontrivial superconductors.  

A correlation function is a key quantity of connecting topological characters with bulk measurements. 
Current-current correlation functions signal the topological invariant in integer Hall systems as a quantized conductance, for example. 
In topological superconductors, the authors proposed that spin-spin correlation functions, measured by the nuclear magnetic relaxation (NMR) rate ($T_{1}^{-1}$), 
can detect their topological nature\cite{PhysRevB.92.180502}. 
The NMR rate in a spin-singlet $s$-wave superconductor is enhanced just below $T_{\rm c}$, owing to the coherence factor\cite{Tinkham}. 
This coherence peak (Hebel-Slichter peak) comes from the formation of $s$-wave-like Cooper pairs\cite{PhysRev.113.1504,PhysRev.125.159}. 
We claimed that an inverse coherence effect is the signature of a 3D odd-parity fully gapped topological superconducting state 
in time-reversal-invariant multiorbital systems; 
the coherence factor contributes to the NMR rates with an opposite sign to that of the conventional $s$-wave states. 

In this paper, we study the temperature dependence of NMR rates to detect a sign of unconventional topological superconductivity in doped topological insulators {\it M}($=$Cu,Nb,Sr)$_{x}$Bi$_{2}$Se$_{3}$ and Sn$_{1-x}$In$_{x}$Te.  
We focus on both isotropic and anisotropic superconducting states. 
Our model is a massive Dirac Hamiltonian with a superconducting gap. 
This is a minimal model of 3D time-reversal-invariant multi-band topological superconductors. 
Moreover, we add a hexagonal warping term to the normal-electron Hamiltonian,  allowing us to argue an effect of a six-fold in-plane\cite{PhysRevB.90.100509}.
We reveal that the NMR rate in a 3D doped topological insulator becomes a tool to detect topologically nontrivial unconventional superconductivity, even with the hexagonal warping term.

This paper organizes as follows. 
Section \ref{sec:model} 
shows our mean-field superconducting model of doped topological insulators. 
We also show the explicit formula of the NMR rate. 
In Sec.~\ref{sec:sec3}, we show the approximate formulation of the NMR rate below $T_{\rm c}$. 
In Sec.~\ref{sec:sec4}, the numerical results are shown. 
We discuss the effect of the hexagonal warping term. 
Section \ref{sec:sec5} shows the discussion. 
The summary is given in Sec.~\ref{sec:sec6}.

\section{Model}
\label{sec:model}
The mean-field Bogoliubov-de Gennes (BdG) Hamiltonian is 
${\cal H} = (1/2) \sum_{\Vec{k}} \Vec{\psi}_{\Vec{k}}^{\dagger} \check{H}(\Vec{k}) \Vec{\psi}_{\Vec{k}}$, 
with $\Vec{\psi}_{\Vec{k}} = (\Vec{c}_{\Vec{k}}^{\dagger},\Vec{c}_{-\Vec{k}}^{\rm T})$ and 
$\Vec{\psi}_{\Vec{k}}^{\rm T} = (\Vec{c}_{\Vec{k}},\Vec{c}_{-\Vec{k}}^{\dagger})$. 
The $2n_{\rm o}$-component column (raw) vector $\Vec{c}_{\Vec{k}}$ ($\Vec{c}_{\Vec{k}}^{\dagger}$) contains electron's annihilation (creation) operators, 
with the number of orbitals $n_{\rm o}$. 
When $n_{\rm o} = 2$, we have $\Vec{c}^{\rm T}_{\Vec{k}} = (c_{\up \Vec{k}}^{1},c_{\down \Vec{k}}^{1},c_{\up \Vec{k}}^{2},c_{\down \Vec{k}}^{2})$. 
The BdG matrix $\check{H}(\Vec{k})$ is 
\begin{align}
\check{H}(\Vec{k}) &= \left(\begin{array}{cc}
\hat{H}_{0}(\Vec{k}) & \hat{\Delta}(\Vec{k}) \\
 \hat{\Delta}^{\dagger}(\Vec{k}) & -\hat{H}_{0}(-\Vec{k})^{\ast}
 \end{array}\right),
\end{align}
where the normal-state Hamiltonian matrix is $\hat{H}_{0}(\Vec{k})$. 
The pairing potential fulfills $\hat{\Delta}^{\rm T}(\Vec{k}) = - \hat{\Delta}(-\Vec{k})$, owing to the Fermion anticommutation property. 
$\check{A}$ signifies the $2 n_{\rm o} \times 2 n_{\rm o}$ matrix structure in the Nambu-Gor'kov particle-hole space, whereas $\hat{A}$ does 
the $n_{\rm o} \times n_{\rm o}$ matrix structure in the orbital-spin space. 

In this paper, we focus on a minimal model for 
3D time-reversal-invariant topological superconductor. 
A typical candidate for a topological superconductor 
is a doped topological insulator with a strong spin orbit coupling, such as $M_x \mbox{Bi}_2 \mbox{Se}_3$ and $\mbox{Sn}_{1-x} \mbox{In}_x \mbox{Te}$. 
The low-energy normal-state $\Vec{k} \cdot \Vec{p}$ Hamiltonian around a time-reversal-invariant point in the momentum space (e.g. $\Gamma$ point in {\it M}$_{x}$Bi$_{2}$Se$_{3}$ or $L$ point in Sn$_{1-x}$In$_{x}$Te ) is given by a massive Dirac Hamiltonian\cite{PhysRevLett.105.097001,PhysRevLett.107.217001,PhysRevB.86.094507,ncomms1969,PhysRevLett.109.217004,JPSJ.83.064703} 
\begin{align}
\hat{H}_{0}(\Vec{k}) 
&= 
\gamma^{0}\bigg[
-\mu \Gamma^{0}
+
\sum_{i=1}^{3} v_{i} k_{i} \Gamma^{i}
+
m \Gamma^{4}
+
h_{5}(\Vec{k})\Gamma^{5}
\bigg], 
\label{eq:normalH}
\end{align}
with chemical potential $\mu$, spin-orbit coupling constants $v_{i}$,
and mass $m$.  
This Hamiltonian is the same as that in Ref.~\onlinecite{PhysRevLett.105.097001} except
$h_{5}$. 
The last term $h_{5}(\Vec{k}) \equiv i \lambda (k_{+}^{3}+k_{-}^{3})$ with $k_{\pm} = k_{x} \pm i k_{y}$ corresponds to the effects of hexagonal warping in the
Fermi surface of $M_{x}\mbox{Bi}_{2}\mbox{Se}_{3}$~\cite{PhysRevB.90.100509}. 
Six kinds of $4\times 4$ matrices $\Gamma^{A}$
($A=0\,,1\,\ldots,5$) are composed of the gamma matrices  
$\gamma^{\mu}$ ($\mu=0,\,1,\,2,\,3$)\cite{Peskin} and the identity: 
\(
\Gamma^{A} = \gamma^{A}
\) 
($A \neq 4$)
and 
\(
\Gamma^{4} = \openone_{4}
\), 
with $\gamma^{5} = i\gamma^{0}\gamma^{1}\gamma^{2}\gamma^{3}$. 
Our choice~\cite{note} is that 
\(
\gamma^{0} = \sigma_{x} \otimes \openone_{2}
\), 
\(
\gamma^{1} = -i \sigma_{y} \otimes s_{y}
\),
\(
\gamma^{2} = i \sigma_{y} \otimes s_{x}
\),
and
\(
\gamma^{3} = i \sigma_{z} \otimes \openone_{2}
\), 
where 
$\sigma_{x,\,y,\,z}$ ($s_{x,\,y,\,z}$) are the $2 \times 2$ Pauli
matrices in the orbital (spin) space. 
In this paper, we study the momentum-independent pair potential $\hat{\Delta}$, owing to 
the onsite interaction~\cite{PhysRevLett.105.097001,PhysRevB.81.220504}.
The fermion anticommutation relation  leads to 
\begin{align}
\hat{\Delta}^{A} 
&= \Delta^{A} \Gamma^{A} \gamma^{2} \gamma^{5}. 
\end{align}
Since $\gamma^{2}\gamma^{5} = \openone_{2}\otimes s_{y}$, the case of
$\Gamma^{A}$ to be the identity (i.e., $A=4$) describes a spin-singlet
$s$-wave state: 
\(
\Delta^{4}
\propto
\sum
\langle 
c_{-\Vec{k} \downarrow}^{1}c_{\Vec{k} \uparrow}^{1}
+
c_{-\Vec{k} \downarrow}^{2}c_{\Vec{k} \uparrow}^{2}
\rangle
\). 
The additional $\Gamma^{A}$ ($A \neq 4$) characterizes 
a \textit{twist} of each order-parameter component in the orbital-spin
space, compared to the conventional $s$-wave state. 
According to Ref.~\onlinecite{PhysRevLett.105.097001}, even-parity order parameters ($A_{1g}$
states) are given by $\Delta^{4}$ and $\Delta^{0}$. 
Odd-parity states correspond to $\Delta^{1,2,3,5}$. 
The component $\Delta^{5}$ corresponds to an odd-parity
fully-gapped ($A_{1u}$) state~\cite{PhysRevLett.105.097001,PhysRevB.81.220504}: 
\(
\Delta^{5}
\propto
\sum 
\langle
c_{-\Vec{k} \downarrow}^{2}c_{\Vec{k} \uparrow}^{1}
+
c_{-\Vec{k} \uparrow}^{2}c_{\Vec{k} \downarrow}^{1}
\rangle
\).
The others are anisotropic odd-parity topological states; $\Delta^{1}$ and $\Delta^{2}$ ($E_{u}$ states) have deep minima, respectively, in the directions of $x$- and $y$- axes, whereas $\Delta^{3}$ ($A_{2u}$ state) does so in $z$ direction. Specifically, the 
odd-parity states with two-fold deep gap minima on $a$-$b$ plane are composed of 
\(
\Delta^{1}
\propto
\sum 
\langle
c_{-\Vec{k} \uparrow}^{1}c_{\Vec{k} \uparrow}^{2}
-
c_{-\Vec{k} \downarrow}^{1}c_{\Vec{k} \downarrow}^{2}
\rangle
\) 
and 
\(
\Delta^{2}
\propto
\sum 
\langle
c_{-\Vec{k} \uparrow}^{1}c_{\Vec{k} \uparrow}^{2}
+
c_{-\Vec{k} \downarrow}^{1}c_{\Vec{k} \downarrow}^{2}
\rangle
\). 
In the case of $E_{g}$ pairing without the hexagonal warping term (e.g.~$\lambda = 0$), the superconducting order parameter 
with point-nodes in $\theta_{\rm N}$ direction is expressed as\cite{JPSJ.83.063705}
\begin{align}
\hat{\Delta}_{\theta_{\rm N}} &= (\cos \theta_{\rm N} \hat{\Delta}^{1} + \sin \theta_{\rm N} \hat{\Delta}^{2}).
\end{align}
The point nodes are located at\cite{JPSJ.83.063705} 
\begin{align}
\Vec{k}_{\rm node}^{\pm} &= \pm \sqrt{\mu^{2} - m^{2} + |\Delta|^{2}} (\cos \theta_{\rm N},\sin \theta_{\rm N},0), 
\end{align}
Here, we adopt $|\Delta|^{2} = |\Delta^{1}|^{2} = |\Delta^{2}|^{2}$.
With increasing $\lambda$, the point-nodes change to the deep minima in the momentum space.\cite{PhysRevB.90.100509}
In this paper, we set $\theta_{\rm N} = 0$; the nodes are lifted up by the hexagonal warping term, resulting in a full gap.

The NMR rate~\cite{PhysRevB.92.180502,PhysRevB.73.092508,NewJPhys.10.103026} in a multi-orbit superconductor is calculated by 
\begin{align}
&\frac{1}{T_{1}(T) T} 
=  
\pi  \sum_{\alpha, \alpha^{\prime}} 
\int_{-\infty}^{\infty} d \omega \,
\left[
- \frac{d f(\omega)}{d \omega} 
\right]
\nonumber \\
& \times {\rm Re} \: 
\left\{ 
\rho_{\up \up}^{G \alpha \alpha^{\prime}}(\omega) 
\rho_{\down \down}^{G  \alpha^{\prime} \alpha}(\omega) 
-
\rho_{\up \down}^{F \alpha \alpha^{\prime}}(\omega) 
[\rho_{\down \up}^{F \alpha \alpha^{\prime}}(\omega)]^{\ast} 
\right\}. \label{eq:t1}
\end{align}
We use the unit system of $\hbar=k_{\rm B}=1$. 
The indices $\alpha$ and $\alpha^{\prime}$ represent orbital labels. 
The Fermi-Dirac distribution function is denoted by $f(\omega) = 1/(e^{\omega/T} + 1)$.
The spectral functions $\hat{\rho}^{G}(\omega)$ and 
$\hat{\rho}^{F}(\omega)$ are the submatrices of 
\begin{align}
\check{\rho}^{G}(\omega) 
= \frac{-1}{2 \pi i}
\sum_{\Vec{k}} 
\left[ 
\check{G}_{\Vec{k}}(i\omega_{n} \rightarrow \omega + i 0) 
- 
\check{G}_{\Vec{k}}(i\omega_{n} \rightarrow \omega - i 0)  
\right],
\end{align}
with the temperature Green's function defined as 
\begin{align}
\check{G}_{\Vec{k}}(i\omega_{n})
&=
[i\omega_{n} - \check{H}(\Vec{k})]^{-1}.
\end{align}
Here, the fermionic Matsubara frequency is $\omega_{n}=\pi T(2n+1)$ 
($n\in \mathbb{Z}$). 
The matrix form of $\check{G}_{\bm k}(i \omega_{n})$ in the Nambu-Gor'kov particle-hole space is 
\begin{align}
 \check{G}_{\Vec{k}}(i\omega_{n}) 
&= 
\left(\begin{array}{cc}
 \hat{G}_{\Vec{k}}(i\omega_{n}) & \hat{F}_{\Vec{k}}(i\omega_{n}) \\
 \hat{\bar{F}}_{\Vec{k}}(i\omega_{n}) &  \hat{\bar{G}}_{\Vec{k}}(i\omega_{n})
\end{array}\right).
\end{align}
The diagonal block, $\hat{G}_{\Vec{k}}$ leads to $\hat{\rho}^{G}$, relevant to electron's density of states.
The off-diagonal block, $\hat{F}_{\Vec{k}}$ contributes to the anomalous spectral function $\hat{\rho}^{F}$. 

\section{Approximate formulation below $T_{\rm c}$} \label{sec:sec3}
A coherence effect just below $T_{\rm c}$ originates from the second term in Eq.~(\ref{eq:t1})~\cite{PhysRevB.73.092508,RevModPhys.63.239}.
The Hebel-Slichter peak appears when this term, including the minus sign in front of the spectral functions, has a positive contribution to $T_{1}^{-1}$. 
To understand the behaviors of the second term, we evaluate anomalous Green's function near $T_{\rm c}$. 
Linearizing $\check{G}$ with respect to $\hat{\Delta}^{A}$, we obtain 
\begin{align}
\hat{F}_{\Vec{k}}^{A}(i \omega_{n}) \approx \hat{G}_{\Vec{k}}^{\rm N}(i \omega_{n}) \hat{\Delta}^{A} \hat{\bar{G}}_{\Vec{k}}^{\rm N}(i \omega_{n}),
\end{align}
with normal-state Green's functions: 
\begin{align}
\hat{G}_{\Vec{k}}^{\rm N}(i \omega_{n}) & \equiv \left[ i \omega_{n} - \hat{H}_{0}(\Vec{k}) \right]^{-1}, \\
 \hat{\bar{G}}_{\Vec{k}}^{\rm N}(i \omega_{n}) &\equiv \left[ i \omega_{n} + \hat{H}_{0}(-\Vec{k})^{\ast} \right]^{-1}.
\end{align}
With the use of the relation $\gamma^{2} \gamma^{5}  \hat{H}_{0}(-\Vec{k})^{\ast} \gamma^{2} \gamma^{5} = \hat{H}_{0}(\Vec{k})$, 
we have 
\begin{align}
 \hat{\bar{G}}_{\Vec{k}}^{\rm N}(i \omega_{n}) &= \gamma^{2} \gamma^{5} (i \omega_{n} + \hat{H}_{0}(\Vec{k}))^{-1} \gamma^{2} \gamma^{5}, \nonumber 
\\
&= -  \gamma^{2} \gamma^{5} \hat{G}_{\Vec{k}}^{\rm N}(-i \omega_{n})  \gamma^{2} \gamma^{5}.
\end{align}
A normal-state Green's function is evaluated by an algebraic relation of $\hat{H}_{0}$; we find that $[\hat{H}_{0}'(\Vec{k})]^{2} = E(\Vec{k})^{2}$, with $\hat{H}_{0}' \equiv 
\hat{H}_{0} + \mu$ and $E(\Vec{k})^{2} = \sum_{i=1}^{3} v_{i}^{2} k_{i}^{2} + m^{2} - h^{5}(\Vec{k})^{2}$. 
This property corresponds to the fact that the Dirac equation is the square root of the Klein-Gordon equation\cite{Peskin}. 
Hence, we
obtain 
\begin{align}
\hat{G}^{\rm N}_{\Vec{k}}( i\omega_{n})
&=
\sum_{\ell=\pm} \frac{\hat{P}_{\ell}(\Vec{k})}{i \omega_{n} - \ell E(\Vec{k}) + \mu},
\end{align}
with the projectors 
\begin{align}
\hat{P}_{\pm}
= 
\frac{1}{2} \left[ 1 \pm \frac{\hat{H}_{0}^{\prime}}{E(\Vec{k})}\right] 
&\equiv \gamma^{0} 
\sum_{A=0}^{5} w_{\ell \Vec{k}}^{A}\Gamma^{A}.
\end{align}
Just below $T_{\rm c}$, the anomalous Green's function $\hat{F}_{\Vec{k}}(i \omega_{n})$ is expressed as 
\begin{align}
\hat{F}_{\Vec{k}}^{A}(i \omega_{n}) &= - \sum_{l,l'=\pm} W_{l l' \Vec{k}}(i \omega_{n}) \hat{P}_{l}(\Vec{k}) \hat{P}_{l'A}(\Vec{k})
\hat{\Delta}^{A} ,
\end{align}
with 
\begin{align}
W_{l l' \Vec{k}}(i \omega_{n}) &\equiv \frac{1}{i \omega_{n} -l E(\Vec{k}) + \mu} \frac{1}{-i \omega_{n} - l' E(\Vec{k}) + \mu}, \\
 \hat{P}_{l'A}(\Vec{k}) &\equiv \Gamma^{A} \hat{P}_{l}(\Vec{k}) [\Gamma^{A}]^{-1} \equiv \gamma^{0} \sum_{A'=0}^{5} w_{l \Vec{k}}^{AA'} \Gamma^{A'}.
\end{align}
Thus, the local anomalous Green's function becomes 
\begin{align}
\sum_{\Vec{k}} F_{\Vec{k}}^{A}(i \omega_{n}) &= \alpha_{A} \hat{\Delta}^{A} + \beta_{A} \gamma^{0} \hat{\Delta}^{A} + \delta_{A} \gamma^{1} \gamma^{5} \hat{\Delta}^{A}.
\label{eq:fsum}
\end{align}
Here, we use the fact that momentum-odd terms in $\hat{P}_{l}(\Vec{k})$ vanish. 
The coefficients $\alpha_{A}$, $\beta_{A}$ and $\delta_{A}$ are defined as 
\begin{align}
\alpha_{A}(i \omega_{n}) &= -\sum_{\Vec{k}} \sum_{l,l=\pm} W_{ll' \Vec{k}}(i \omega_{n}) \sum_{A'=0}^{5} w_{l \Vec{k}}^{A'}w_{l' \Vec{k}}^{AA'}(1-2\delta_{A'5}),\\
\beta_{A}(i \omega_{n}) &=-\sum_{\Vec{k}} \sum_{l,l=\pm} W_{ll' \Vec{k}}(i \omega_{n}) \left( w_{l \Vec{k}}^{4}w_{l' \Vec{k}}^{A0} + w_{l \Vec{k}}^{0}w_{l' \Vec{k}}^{A4}\right),\\
\delta_{A}(i \omega_{n}) &= -\sum_{\Vec{k}} \sum_{l,l=\pm} W_{ll' \Vec{k}}(i \omega_{n}) \left( w_{l \Vec{k}}^{5}w_{l' \Vec{k}}^{A1} - w_{l \Vec{k}}^{1}w_{l' \Vec{k}}^{A5}\right).
\end{align}

The sign of the coherence effect of $T_{1}^{-1}$ is determined by the spin parity of the local anomalous Green's function\cite{PhysRevB.92.180502}.
The first and second terms of $\sum_{\Vec{k}} F_{\Vec{k}}^{A}(i \omega_{n})$ in Eq.~(\ref{eq:fsum}) have a same spin-parity of the order parameter $\hat{\Delta}^{A}$, since 
the multiplication of $\gamma^{0}$ with $\hat{\Delta}^{A}$ does not change the property of a spin-index exchange. 
The third term, which is proportional to the warping term $h_{5}$, rotates spins, since $\gamma^{1} \gamma^{5} = \openone_{2} \otimes s_{x}$. 
For example, in the case of the fully-gapped odd-parity state $\hat{\Delta}^{5} \propto \sigma_{y} \otimes s_{x}$, 
the first and second terms contribute to the anomalous spectral function $\rho_{\up \down}^{\alpha \alpha'} = \rho_{\down \up}^{\alpha \alpha'}$.
The third term does not contribute to the anomalous spectral function, since this term is proportional to $\gamma^{1} \gamma^{5} \hat{\Delta}^{5} \propto \sigma_{y} \otimes \openone_{2}$.
Thus, 
the inverse coherence effect\cite{PhysRevB.92.180502} can appear, irrespective of hexagonal warping. 
In the case of the anisotropic odd-parity state $\hat{\Delta}^{1} \propto \sigma_{y} \otimes \openone_{2}$, 
the spin-singlet element of the gap function $\Delta_{\up \down}^{\alpha \alpha'}$ is zero so that the first and second terms do not contribute to the 
NMR rate. 
The only third term contributes to $\rho_{\up \down}^{\alpha \alpha'} = \rho_{\down \up}^{\alpha \alpha'}$ 
and the amplitude of the inverse coherence effect is proportional to the magnitude of the warping term. 

\section{Numerical results}\label{sec:sec4}
In this section, we show the temperature dependence of the NMR rate $T_{1}^{-1}$ with various odd-parity superconducting phases. 
We assume the phenomenological temperature dependence of the gap amplitude as\cite{NewJPhys.10.103026} 
\begin{align}
\Delta(T) &=\Delta_{0} \tanh \left(a \sqrt{T_{\rm c}/T -1} \right), \label{eq:gapbcs}
\end{align}
with $T_{\rm c} = 1.76 \Delta_{0}$. 
Equation (\ref{eq:gapbcs}) with $a = 1.74$ reproduces well the temperature dependence of the BCS gap. 
We set the gap amplitude $\Delta_{0} = 0.1$, the Dirac mass $m = 0.4$, and the spin-orbit interaction $v_{i} = 1$. 
The $\Vec{k}$ integrals are performed by the trapezoidal rule in the spherical coordinate system, with cutoff momentum $k_{\rm max} = 9$ and mesh 
$(N_{k},N_{\phi},N_{\theta}) = (384,96,96)$. 
The smearing factor of the delta function is set by 0.01. 
To compare with the previous results by self-consistently solving the gap equations\cite{PhysRevB.92.180502}, we set same parameter with $\mu = 0.8$ as that in Ref.~\onlinecite{PhysRevB.92.180502}. 
We introduce the effective gap function on the Fermi surface to discuss the quasiparticle excitations due to the thermal effect. 
According to Eq.(9) in Ref.~\onlinecite{PhysRevB.90.100509}, 
the spectral gap on the Fermi surface in the presence of the hexagonal warping in the $\Delta^{1}$ state is expressed as 
\begin{align}
\Delta^{{\rm spec},1 }(\Vec{k}) = |\Delta| \sqrt{1 - [\tilde{\Vec{k}} \cdot (\hat{\Vec{z}} \times \Vec{n})]^{2}},
\label{eq:deltaspec}
\end{align}
with $\tilde{\Vec{k}} \equiv (k_{x},k_{y},k_{z})/\sqrt{\mu^{2}-m^{2}}$ and $\Vec{n} = (0,1,0)$.

\subsection{Without the warping term $h_{5}$}
\begin{figure}[t]
%%%%--- I comment out figure regions
%\vspace{50mm}
\begin{center}
     \begin{tabular}{p{  \columnwidth}}% p{0.8 \columnwidth}}%  p{28mm}}
      (a) \: \: \resizebox{ \columnwidth}{!}{\includegraphics{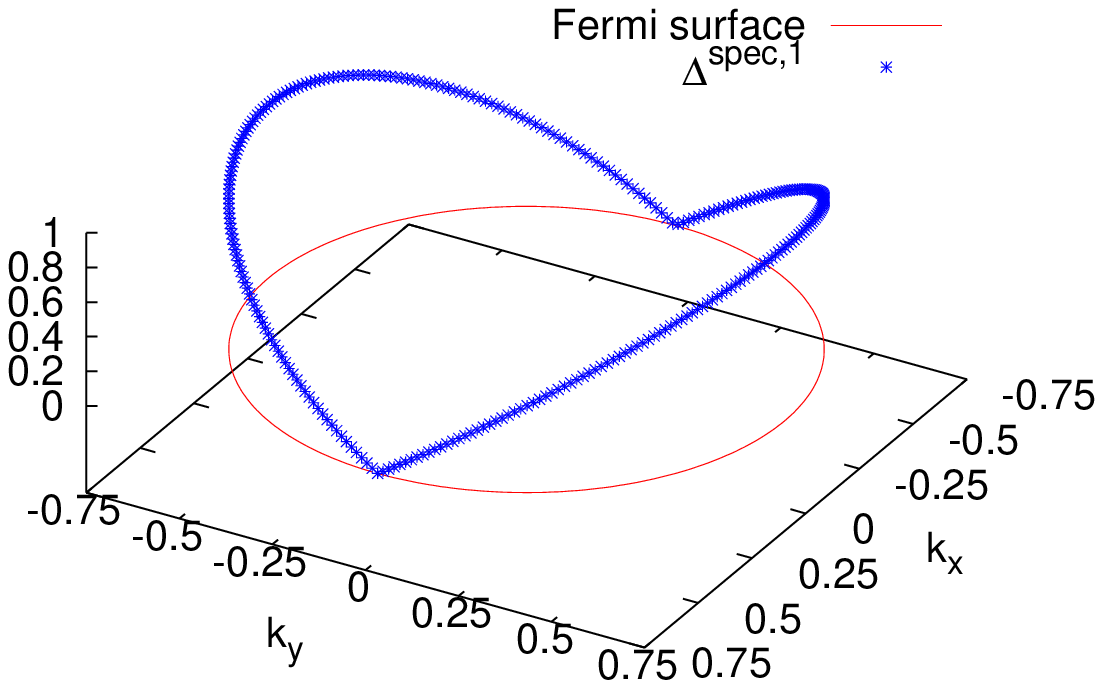}} \\
      (b) \: \:  \resizebox{ \columnwidth}{!}{ \includegraphics{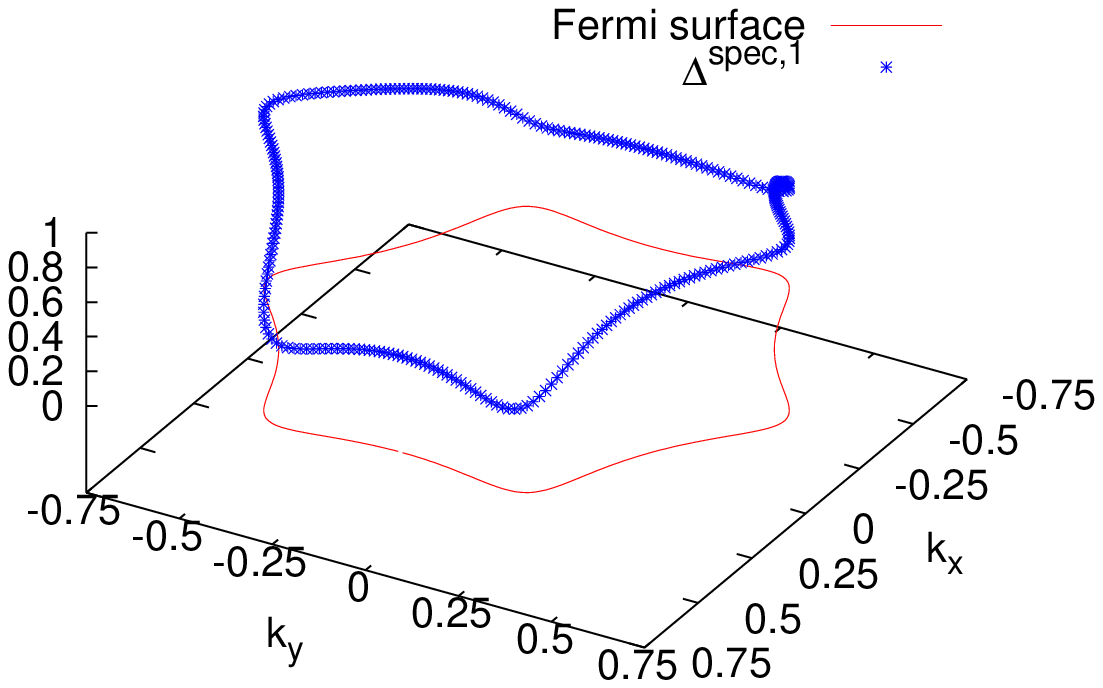}} 
%      \\ %&
%      (c) \: \: \resizebox{0.8 \columnwidth}{!}{\includegraphics{m04mu08delta3_160407.eps}} 
%      & (d) \: \: \resizebox{0.8 \columnwidth}{!}{\includegraphics{m04mu08delta4_160407.eps}} 
    \end{tabular}    
\end{center}
\caption{
\label{fig:Fs}(Color online) The Fermi surface and the spectral gap $\Delta^{{\rm spec},1 }(\Vec{k})$ defined in Eq.~(\ref{eq:deltaspec}) with $k_{z} = 0$ (a) without and (b) with the hexagonal warping term ($\lambda = 1$) to treat the six-fold crystal structure. 
 }
\end{figure}

In this section, we drop the warping term $h_{5}$. 
The Fermi surface in normal states is isotropic as shown in Fig.~\ref{fig:Fs}(a). 
Dropping the warping term, there are point nodes in $x$-direction in the $\Delta^{1}$ state.
The results are summarized as Table \ref{table:1}.

\begin{figure*}[t]
%%%%--- I comment out figure regions
%\vspace{50mm}
\begin{center}
     \begin{tabular}{p{ 0.8 \columnwidth} p{0.8 \columnwidth}}%  p{28mm}}
      (a) \: \: \resizebox{0.8 \columnwidth}{!}{\includegraphics{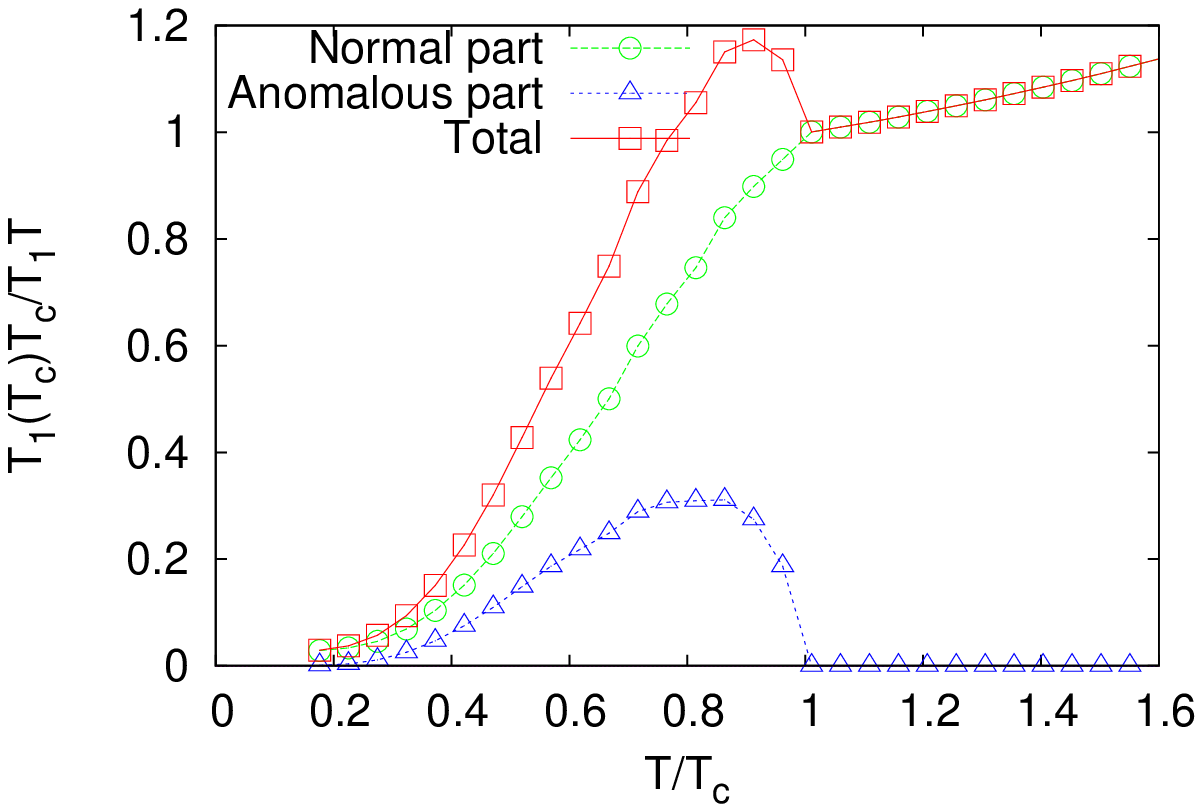}} 
      &(b) \: \:  \resizebox{0.8 \columnwidth}{!}{ \includegraphics{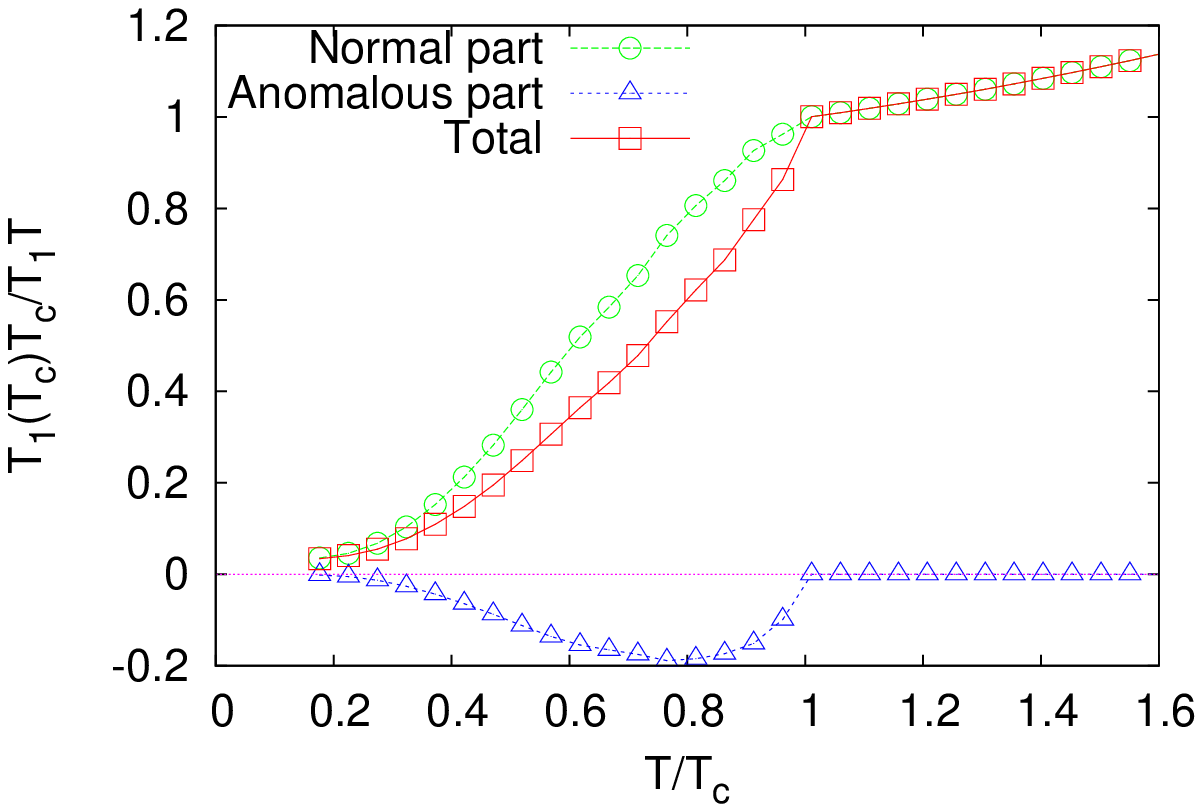}} 
      \\ %&
      (c) \: \: \resizebox{0.8 \columnwidth}{!}{\includegraphics{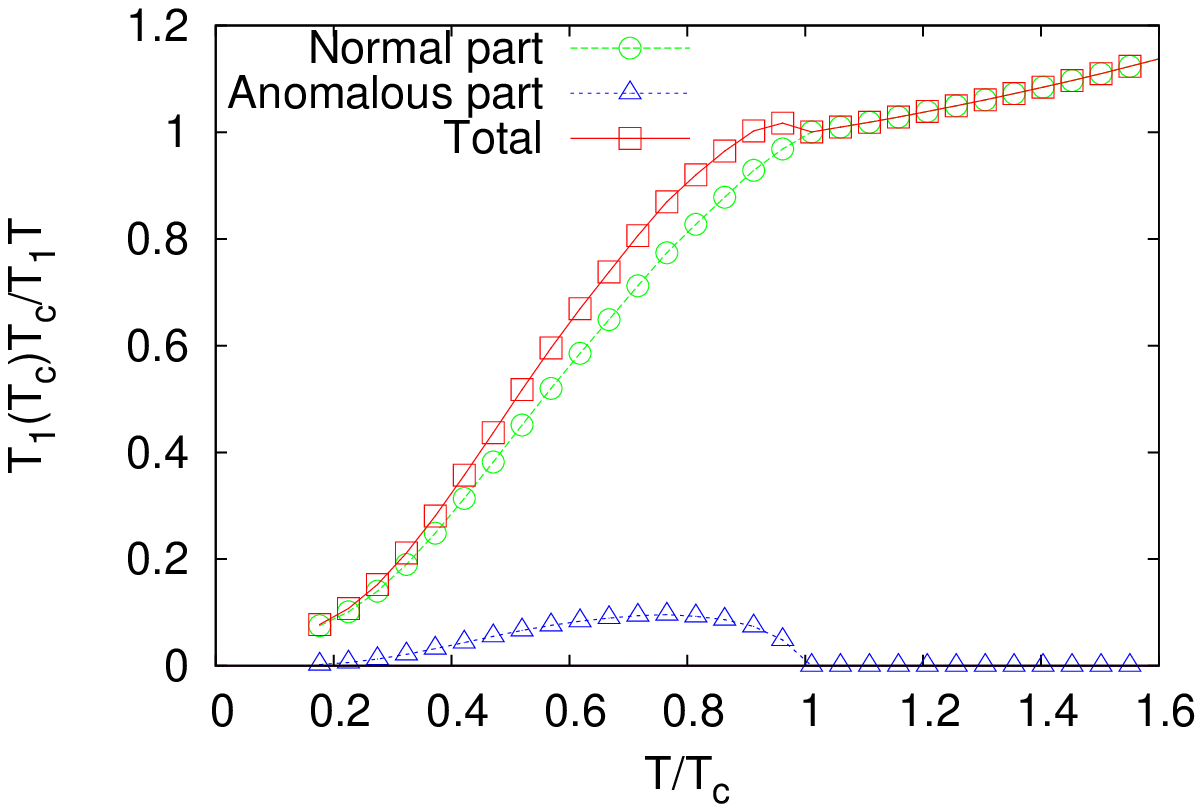}} 
      & (d) \: \: \resizebox{0.8 \columnwidth}{!}{\includegraphics{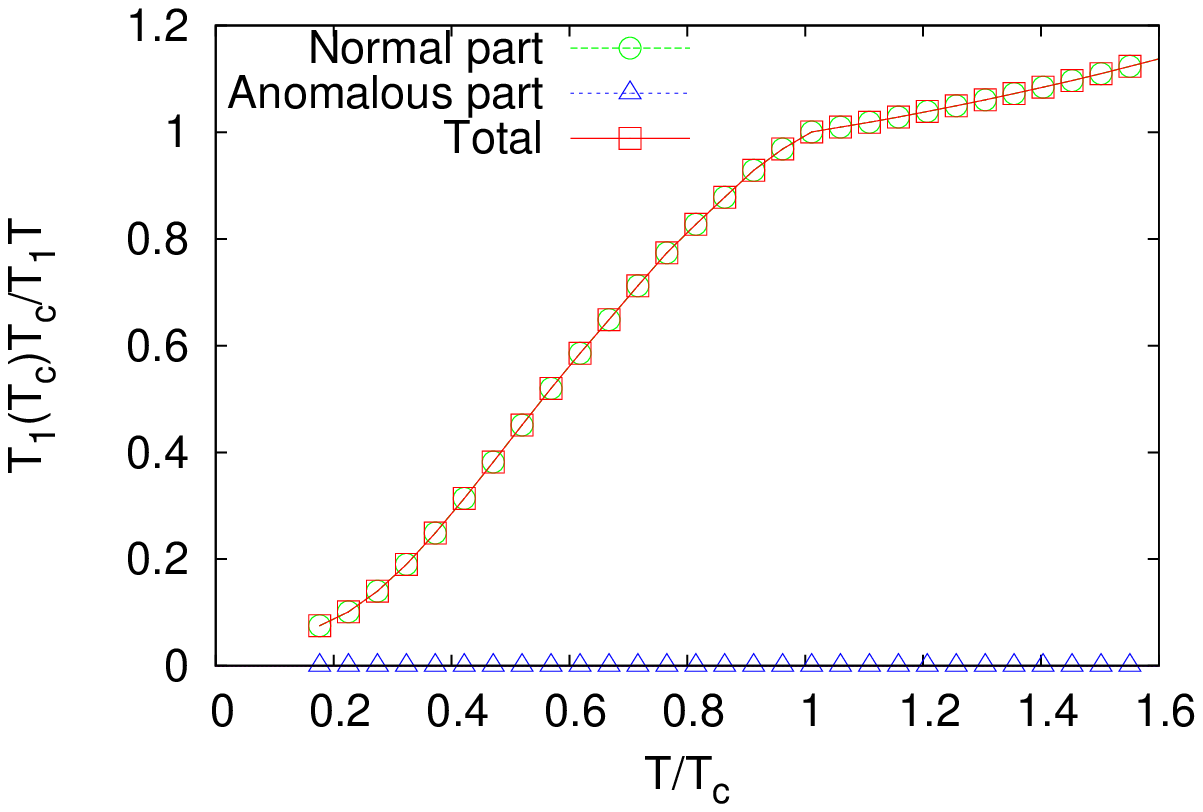}} 
    \end{tabular}    
\end{center}
\caption{
\label{fig:fig2}(Color online) 
Temperature dependence of nuclear magnetic relaxation rates in (a) 
an even-parity gap $\Delta^{4}$ $A_{1g}$)
\(
\propto
\sum
\langle 
c_{-\Vec{k} \downarrow}^{1}c_{\Vec{k} \uparrow}^{1}
+
c_{-\Vec{k} \downarrow}^{2}c_{\Vec{k} \uparrow}^{2}
\rangle
\)
, a fully-gapped isotropic odd-parity gap $\Delta^{5}$ ($A_{1u}$)
\(
\propto
\sum 
\langle
c_{-\Vec{k} \downarrow}^{2}c_{\Vec{k} \uparrow}^{1}
+
c_{-\Vec{k} \uparrow}^{2}c_{\Vec{k} \downarrow}^{1}
\rangle
\)
, 
(c) a nodal odd-parity gap $\Delta^{3}$ ($A_{2u}$)
\(
\propto
\sum
\langle 
c_{-\Vec{k} \downarrow}^{1}c_{\Vec{k} \uparrow}^{1}
-
c_{-\Vec{k} \downarrow}^{2}c_{\Vec{k} \uparrow}^{2}
\rangle
\)
, and (d) a nodal odd-parity gap $\Delta^{1}$ ($E_{u}$)
\(
\propto
\sum
\langle 
c_{-\Vec{k} \uparrow}^{1}c_{\Vec{k} \uparrow}^{2}
-
c_{-\Vec{k} \downarrow}^{1}c_{\Vec{k} \downarrow}^{2}
\rangle
\)
. 
We set the chemical potential $\mu = 0.8$, the Dirac mass $m=0.4$ and the gap amplitude $\Delta_{0} = 0.01$. 
We ignore the warping term.
Equation (\ref{eq:gapbcs}) is used as the phenomenological temperature dependence of the gap amplitude. } 
\end{figure*}

Figure \ref{fig:fig2} shows the temperature dependence of the NMR rate with various kinds of gap functions. 
First, let us argue the results of the isotropic gaps [Figs.~\ref{fig:fig2}(a,b)]. 
Just below $T_{\rm c}$ we confirm the coherence effects predicted in Sec.~III. 
We find a standard behavior at low-temperature region, caused by a full gap. 
We stress that all of the non self-consistent temperature dependences have a good qualitative agreement with those in our previous self-consistent calculations\cite{PhysRevB.92.180502}. 
It indicates that the usage of Eq.~(\ref{eq:gapbcs}) is adequate for investigating the temperature dependence of the NMR rates in doped topological insulators.

\begin{table*}[t]
\caption{
Parity table of different gap functions.
The coherence effect  is characterized by 
spin parity $p_{\rm s}$ [$\Delta_{\up \down}^{\alpha \alpha^{\prime}}(\Vec{k}) = p_{\rm s} \Delta_{\down \up}^{\alpha \alpha^{\prime}}(\Vec{k})$], 
momentum parity $p_{\rm m}$ [$\Delta_{\up \down}^{\alpha \alpha^{\prime}}(\Vec{k}) = p_{\rm m} \Delta_{\up \down}^{\alpha \alpha^{\prime}}(-\Vec{k})$], and 
orbital parity $p_{\rm o}$ [$\Delta_{\up \down}^{\alpha \alpha^{\prime}}(\Vec{k}) = p_{\rm o} \Delta_{\up \down}^{\alpha^{\prime} \alpha}(\Vec{k})$], with $p_{\rm s} p_{\rm m} p_{\rm o} = -1$.  
}
\label{table:1}
\begin{ruledtabular}
\begin{tabular}{ccccccc}
Gap type &
Spin parity $p_{\rm s}$&
Orbital parity $p_{\rm o}$& 
Anisotropic direction &
Coherence effect \\
$A_{1g}(\Delta^{4})$ state & -1 & 
+1 &
Isotropic &
positive \\
$A_{1u}(\Delta^{5})$ state & +1 & 
-1 &
Isotropic &
negative \\
$A_{2u}(\Delta^{3})$ state & -1 & 
+1 &
$z$ &
positive \\
$E_{u}$($\Delta^{1}$ or $\Delta^{2}$) state & +1 & 
-1 &
$x$ or $y$ &
negligible negative 
\end{tabular}
\end{ruledtabular} 
\end{table*}

Next we focus on the anisotropic topological states. 
As shown in Fig.~\ref{fig:fig2}(c), the positive coherence effect appears in the $\hat{\Delta}^{3}$ state, since the spin-parity of the gap function $\hat{\Delta}^{3} \propto \sigma_{z} \otimes s_{y}$ is odd so that the anomalous spectral function becomes $\rho_{\up \down}^{\alpha \alpha'} = -\rho_{\down \up}^{\alpha \alpha'}$. 
In the case of the $\hat{\Delta}^{1}$ state, as shown in Fig.~\ref{fig:fig2}(d), no coherence effect occurs. 
As discussed in Sec.~\ref{sec:sec3}, this gap function without the warping term $h_{5}$ does not have spin-off-diagonal elements of the anomalous spectral function (i.e. $\rho_{\up \down}^{\alpha \alpha'} = \rho_{\down \up}^{\alpha \alpha'} = 0$). 
In low-temperature region ($T \sim 0.2T_{\rm c}$), 
there are low-energy quasiparticle excitations in Fig.~\ref{fig:fig2}(c) and (d), since $\hat{\Delta}^{3}$ ($A_{2u}$) and $\hat{\Delta}^{1}$ ($E_{u}$) states have point-nodes in momentum space. 

\subsection{With the warping term $h_{5}$}
We take the warping term $h_{5}$ into account to consider a six-fold symmetry due to the crystal structure. 
We consider $\lambda = 1$ to describe an anisotropic Fermi surface. 
The Fermi surface with the warping term is shown in Fig.~\ref{fig:Fs}(b). 
On this Fermi surface, the spectral  gap function $\Delta^{{\rm spec},1}(\Vec{k})$ does not have point nodes. 

\begin{figure}[t]
%%%%--- I comment out figure regions
%\vspace{50mm}
\begin{center}
     \begin{tabular}{p{ 0.8 \columnwidth}}% p{0.8 \columnwidth}}%  p{28mm}}
      (a) \: \: \resizebox{ 0.8 \columnwidth}{!}{\includegraphics{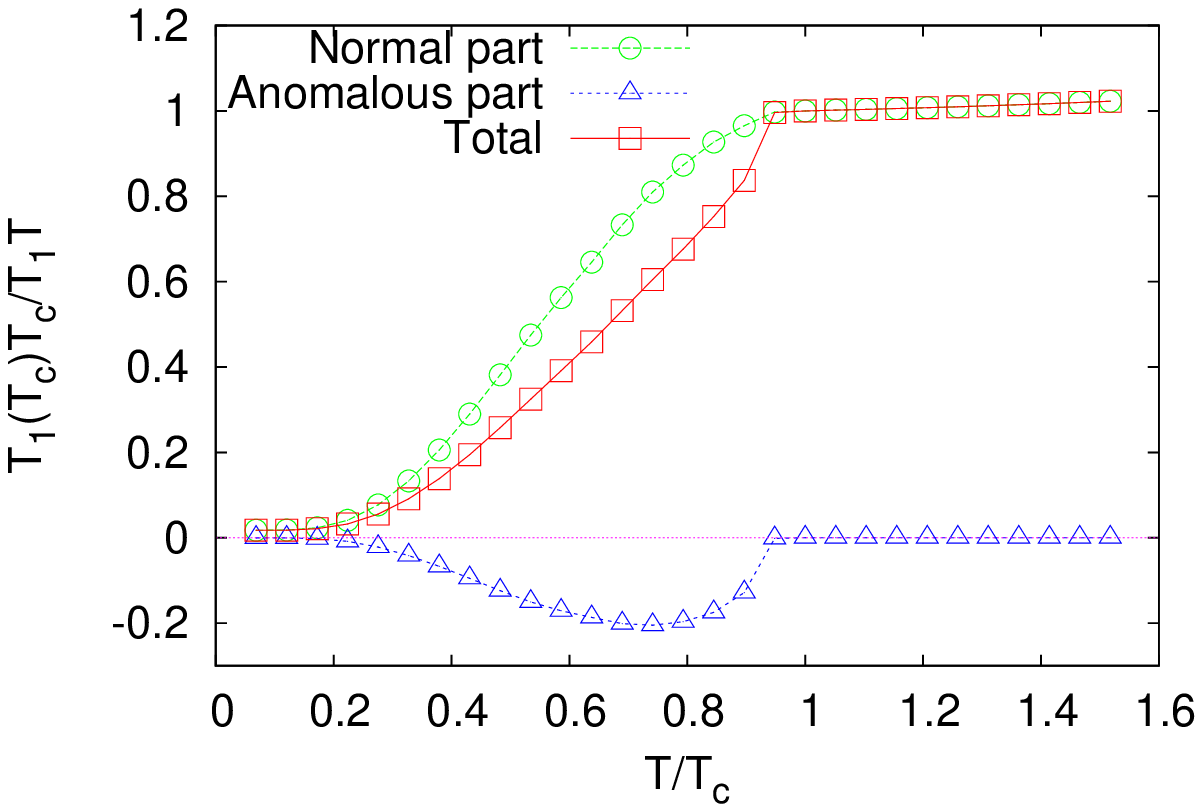}} \\
      (b) \: \:  \resizebox{ 0.8 \columnwidth}{!}{ \includegraphics{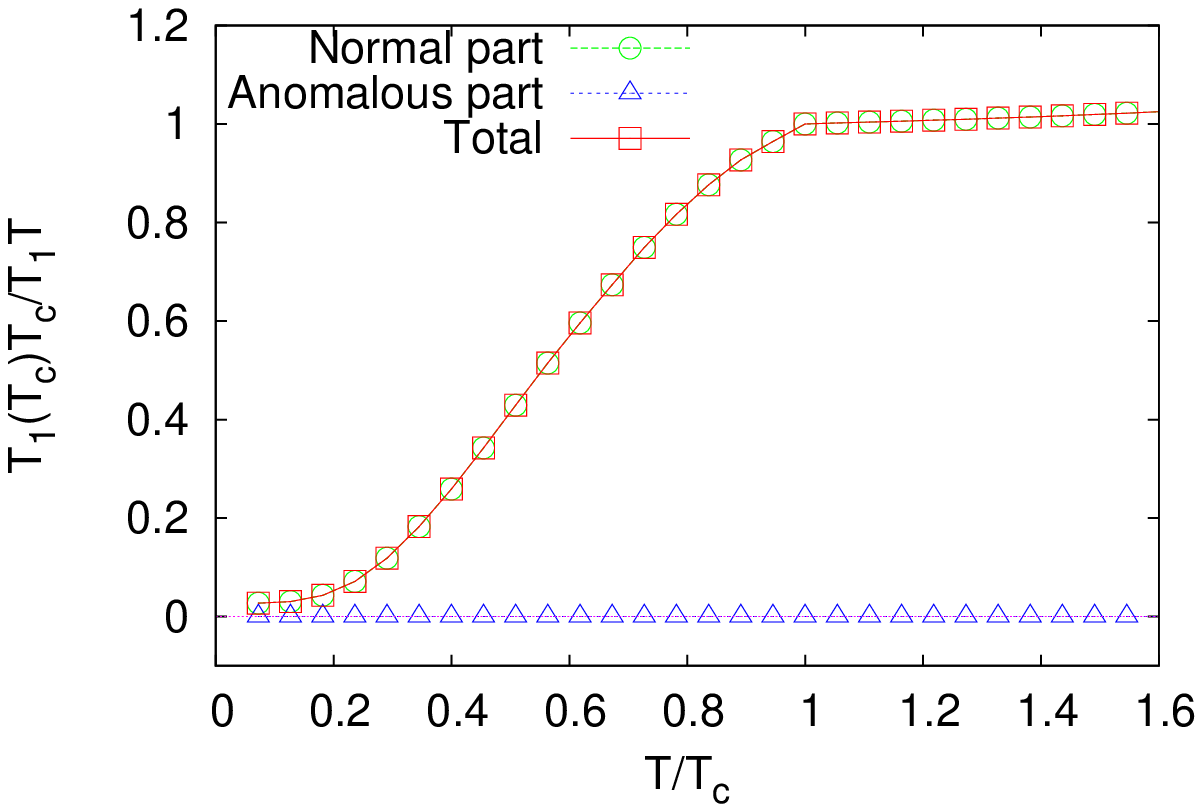}} 
%      \\ %&
%      (c) \: \: \resizebox{0.8 \columnwidth}{!}{\includegraphics{m04mu08delta3_160407.eps}} 
%      & (d) \: \: \resizebox{0.8 \columnwidth}{!}{\includegraphics{m04mu08delta4_160407.eps}} 
    \end{tabular}    
\end{center}
\caption{
\label{fig:fig3}(Color online) 
Temperature dependence of nuclear magnetic relaxation rates with the hexagonal warping term in (a) a fully-gapped isotropic odd-parity gap $\Delta^{5}$ ($A_{1u}$) and (b) an anisotropic odd-parity gap $\Delta^{1}$ ($E_{u}$). The parameters are same in Fig.~\ref{fig:fig2}
 }
\end{figure}

In both $\Delta^{5}$($A_{1u}$) and $\Delta^{1}$($E_{u}$) states, the warping term slightly changes the temperature dependence of the NMR rate, as shown in Fig.~\ref{fig:fig3}. 
The difference between Fig.~\ref{fig:fig2}(a) and Fig.~\ref{fig:fig3}(a) with the $\Delta^{5}$ gap function comes from the density of states on the Fermi surface, since 
the third term in Eq.~(\ref{eq:fsum}), induced by the warping term, does not contribute to the NMR rate. 
In the case of the $\Delta^{1}$ gap function, an inverse coherence effect induced by the warping term may occur, according to Sec.~\ref{sec:sec3}. 
The numerical calculation can conclude, however, that the induced coherence effect is negligible small as shown in Fig.~\ref{fig:fig3}(b). 
We confirm that the relativistic indicator, introduced in Ref.~\onlinecite{PhysRevB.92.180502}, does not affect the induced coherence effect, by changing the Dirac mass $m$ and the chemical potential $\mu$.  
This result comes from the fact that the warping term is the third order of the momentum so that the summation in whole momentum space becomes small. 

At the low temperature region ($T < 0.2 T_{\rm c}$), the amplitude of $1/T_{1}T$ in Fig.~\ref{fig:fig3}(b) is smaller than that in Fig.~\ref{fig:fig2}(d). 
This originates from the fact that there is no point-node in the $\Delta^{1}$ state with the warping term as shown in Fig.~\ref{fig:Fs}(b). 
Note that this difference might be too small to identify whether there are point-nodes or not at low temperatures. 
By comparison with the numerical results with and without the warping term, we conclude that the warping term does not affect the temperature dependence of the NMR rate near $T_{\rm c}$.

\section{Discussion}\label{sec:sec5}
We discuss the amplitude of the coherence effects.
As we discussed in Ref.~\onlinecite{PhysRevB.92.180502}, the relativistic indicator defined by 
$\beta \equiv \sqrt{(\mu/m)^{2}-1}$ characterizes the amplitude of the coherence factor. 
This indicator is controlled by the chemical potential shift. 
With increasing the chemical potential $\mu$, the indicator $\beta$ increases. 
In the ultrarelativistic limit $\beta \rightarrow \infty$ (i.e. $\mu \rightarrow \infty$ or $m \rightarrow 0$), the amplitude of the coherence effect becomes largest in the fully-gapped isotropic topological superconducting state $\Delta^{5}$ ($A_{1u}$), as shown in Fig.~3 in Ref.~\onlinecite{PhysRevB.92.180502}. 
We confirm that the similar behaviors occur in other topologically nontrivial superconducting states $A_{2u}$ and $E_{u}$. 

The robustness of the coherence effect against impurities is important information to measure the NMR rate in experiments. 
We discussed the robustness of the density of states against impurities in Refs. \onlinecite{PhysRevB.89.214506,PhysRevB.91.060502}. 
We concluded that the relativistic indicator $\beta$ characterizes the impurity effect. 
In the nonrelativistic limit $\beta \rightarrow 0$ (i.e. $\mu \rightarrow m$), 
the topological superconducting states $A_{1u}$ and $E_{u}$ are fragile against nonmagnetic impurities, since 
the effective gap functions are $p$-wave ones\cite{PhysRevB.91.060502,JPSJ.83.053705}. 
In this limit, the Dirac-BdG Hamiltonian is regarded to the BdG Hamiltonian. 
As we discussed above, the amplitude of the coherence effects in these topological states is proportional to the indicator $\beta$ so that one does not observe the coherence effects below $T_{\rm c}$ even without impurities. 
On the other hand, the low-energy density of states is robust against nonmagnetic impurities in the ultrarelativistic limit. 
In this limit, the Dirac-BdG equations are divided into a left-handed sector and a right-handed sector as discussed in Ref.~\onlinecite{PhysRevB.92.180502}. 
The effective gap functions are $s$-wave ones. 
Thus, the large amplitude of the coherence effect is robust against nonmagnetic impurities. 
We reveal that the NMR rate in a 3D doped topological insulator becomes a tool to detect topologically nontrivial unconventional superconductivity.

\section{Summary}\label{sec:sec6}
In conclusion, we studied the temperature dependence of the NMR rate in topologically nontrivial superconducting states in doped topological insulators. 
We found that an inverse coherence effect occurs in a fully-gapped isotropic odd-parity state and a negligible small inverse coherence effect occurs in a strong in-plane anisotropic odd-parity state. 
The hexagonal warping term to describe the six-fold crystal structure does not affect the temperature dependence of the NMR rate near $T_{\rm c}$.
At the low temperature region ($T < 0.2 T_{\rm c}$), the amplitude of $1/T_{1}T$ with the warping term is smaller than that without the warping term. 
However, this difference might be too small to identify whether there are point-nodes or not at low temperatures. 
We reveal that the NMR rate in a 3D doped topological insulator becomes a tool to detect topologically nontrivial unconventional superconductivity.

\section*{Acknowledgment}
The authors would like to acknowledge Hiroki Nakamura and Masahiko Machida for helpful discussions and comments. 
The calculations were performed by the supercomputing system SGI ICE X at the Japan Atomic Energy Agency. 
This study was partially supported by JSPS KAKENHI Grant Number 26800197 and the “Topological Materials Science” (No. 16H00995) KAKENHI on Innovative Areas from JSPS of Japan.

%\section*{references}

\pagebreak

\clearpage
%\newpage
%\widetext
%\onecolumngrid
\setcounter{equation}{0}

\renewcommand{\thefigure}{S\arabic{figure}} 

\setcounter{figure}{0}

\renewcommand{\thesection}{S\arabic{section}.}


\begin{thebibliography}{99}
\bibitem{note}
Our choice is related to the Dirac representation $\gamma^{i}_{\rm D}$ by 
$\gamma^{i} = \hat{P} \gamma^{i}_{\rm D} \hat{P}^{\dagger}$ with $\hat{P}_{11} = s_{0}$, $\hat{P}_{12} = -i s_{z}$,$\hat{P}_{21} = s_{0}$, and $\hat{P}_{22} = i s_{z}$. $s_{0}$ is a unit matrix in spin space.
\end{thebibliography}
\end{document}